\documentclass{article}

 \usepackage[preprint]{neurips_2025}

\usepackage{titlesec}
\titlespacing*{\section}
  {0pt}    
  {6pt}   
  {0pt}    

\usepackage[utf8]{inputenc} 
\usepackage[T1]{fontenc}    
\usepackage{hyperref}       
\usepackage{url}            
\usepackage{booktabs}       
\usepackage{amsfonts}       
\usepackage{nicefrac}       
\usepackage{microtype}      
\usepackage{xcolor}         
\usepackage{graphicx}
\usepackage[most]{tcolorbox}
\usepackage{titlesec}
\titlespacing*{\section}{0pt}{0.2ex}{0.2ex}  
\titlespacing*{\subsection}{0pt}{0.2ex}{0.2ex}
\titlespacing*{\subsubsection}{0pt}{0.2ex}{0.2ex}
\tcbuselibrary{listingsutf8}
\definecolor{airforceblue}{rgb}{0.36, 0.54, 0.66}
\definecolor{bluegray}{rgb}{0.4, 0.6, 0.8}
\definecolor{bleudefrance}{rgb}{0.19, 0.55, 0.91}
\hypersetup{colorlinks,linkcolor={airforceblue},citecolor={bleudefrance},urlcolor={bluegray}}
\definecolor{data}{HTML}{538EA6}
\definecolor{jir}{HTML}{D94A56}
\definecolor{understand}{HTML}{8CBEB2}
\definecolor{explore}{HTML}{F3B562}
\definecolor{incoming}{rgb}{0.271, 0.769, 0.690}
\definecolor{xx}{rgb}{0.851, 0.196, 0.196}
\usepackage{xspace}
\usepackage{multirow}
\usepackage{booktabs}
\newcommand{\jirarena}{\textsc{JIR-Arena}\xspace}
\newcommand{\gpt}{\textsc{gpt-4o}\xspace}
\newcommand{\ds}{\textsc{DeepSeek-V3-0324}\xspace}
\newcommand{\claude}{\textsc{claude-3-7}\xspace}
\newcommand{\gemini}{\textsc{gemini-2.0-flash}\xspace}
\newcommand{\st}{\textsc{sentence-transformers/all-MiniLM-L6-v2}\xspace}

\NewDocumentCommand{\todo}{ mO{} }{\textcolor{incoming}{\textsuperscript{\textit{TODO}}\textsf{\textbf{\small[#1]}}}}
\NewDocumentCommand{\ke}{ mO{} }{\textcolor{cyan}{\textsuperscript{\textit{Ke}}\textsf{\textbf{\small[#1]}}}}
\NewDocumentCommand{\cheng}{ mO{} }{\textcolor{blue}{\textsuperscript{\textit{Cheng}}\textsf{\textbf{\small[#1]}}}}

\newtcolorbox{prompt}[2][]{%
  colback=data!2!white,
  colframe=data!85!black,
  fonttitle=\small,
  title={\texttt{#2}},
  #1
}

\title{JIR-Arena: The First Benchmark Dataset for Just-in-time Information Recommendation}

\author{%
  Ke Yang, Kevin Ros, Shankar Kumar Senthil Kumar, ChengXiang Zhai\\
  University of Illinois Urbana-Champaign \\
  \texttt{\{key4, kjros2, sks10, czhai\}@illinois.edu} \\
}

\begin{document}

\maketitle

\begin{abstract}
Just-in-time Information Recommendation (JIR) is a service that delivers the most relevant information precisely when users need it the most. It plays a critical role in filling users' information gaps during pivotal moments like those in learning, work, and social interactions, thereby enhancing decision-making quality and life efficiency with minimal user effort. Recent device-efficient deployment of performant foundation models and the proliferation of intelligent wearable devices have made the realization of always-on JIR assistants feasible. However, despite the potential of JIR systems to transform our daily life, there has been no prior systematic effort to formally define JIR tasks or establish evaluation frameworks. To bridge this gap, we present the first comprehensive mathematical definition of JIR tasks and their associated evaluation metrics. Furthermore, we introduce \jirarena, the first multimodal JIR benchmark dataset with diverse and information-request-intensive scenarios, designed to evaluate JIR systems across multiple dimensions, including whether they can \textit{i)} accurately infer user information needs, \textit{ii)} provide timely and helpfully relevant recommendations, and \textit{iii)} effectively avoid the inclusion of irrelevant content that might distract users. 

Constructing a JIR benchmark dataset is challenging due to the subjectivity limitations of any individual in estimating the user information need distribution and the uncontrollable JIR systems' variables that hinder reproducible evaluation. To address these, \jirarena: \textit{i)} combines multiple humans and large AI models to approximate the information need distribution; \textit{ii)} evaluates JIR instances' quality based on the information retrieval outcome using static snapshots of knowledge bases; and \textit{iii)} employs a multi-turn, multi-entity validation framework to enhance \jirarena's objectivity and generality. Additionally, we implement a baseline JIR system that processes sensory information streams consistent with user inputs and provides real-time JIR instances. Our evaluation of the baseline on \jirarena reveals that while large foundation model-based JIR systems can simulate user needs with reasonable precision, they struggle with recall and effective content retrieval. Finally, to facilitate future development of JIR systems and exploration of more JIR application scenarios, we fully release our \href{https://github.com/EmpathYang/JIR-Arena}{code} and \href{https://huggingface.co/datasets/EmpathYang/JIR-Arena}{data}.
\end{abstract}

\section{Introduction}
\begin{figure*}[ht]
    \centering
    \includegraphics[trim={0 0.07cm 0 0}, width=1.0\textwidth]{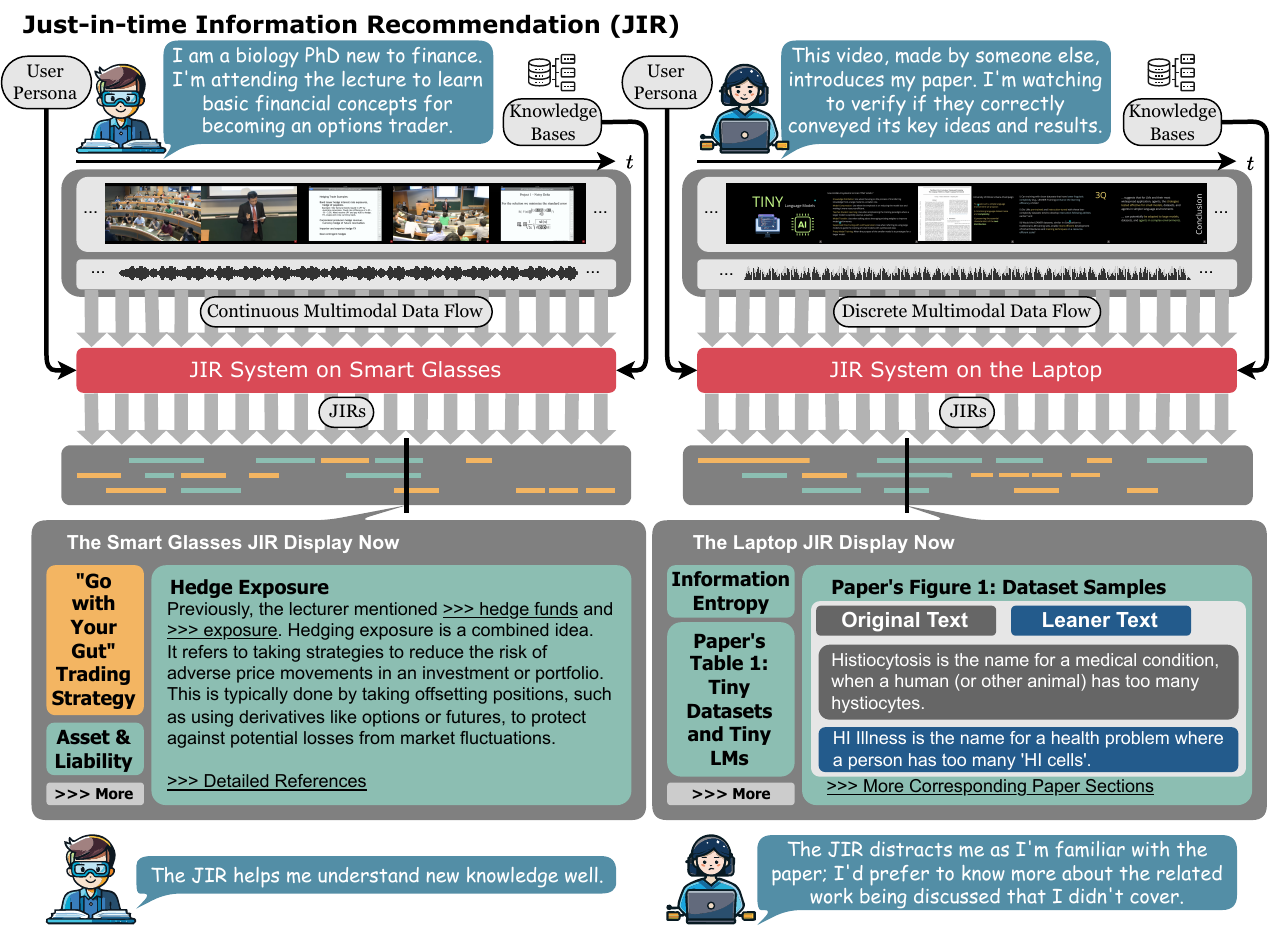}
    \caption{\label{fig:jir_system}
    The JIR system takes in user persona and multimodal data stream, and proactively outputs JIRs based on knowledge bases to satisfy the user's instant information needs.
    }
\end{figure*}
Just-in-time Information Recommendation (JIR) proactively delivers contextually relevant information when needed, enhancing users' productivity and experience. For instance, upon entering a new building, JIR can display floor plans with your location for efficient navigation. Or as shown in Figure \ref{fig:jir_system}, during lectures it can instantly provide succinct explanations of unfamiliar concepts to help you catch up. Promisingly, the recent convergence of portable/wearable devices \citep{meizu_smart_glass,meta_smart_glass} and affordable AI deployment \citep{microsoft2025phi4minitechnicalreportcompact,allal2025smollm2smolgoesbig} has made possible performant AI-powered JIR systems to be integrated into everyday devices—from smartphones to VR glasses—capable of addressing immediate information needs without unnecessary distractions.

As a new paradigm of information service where AI-powered agents can be naturally deployed, JIR will transform how people access information in the future, potentially eliminating their cognitive efforts to find and evaluate relevant information. Despite its significance, the previous closely related concept to JIR, known as just-in-time information retrieval (JITIR), is designed primarily as an information retrieval task and introduced over two decades ago \citep{rhodes2000just}. And unfortunately, progress on JITIR has been limited since its inception due to many technical limitations (see Section~\ref{app:related_work} for further discussion). Therefore, prior work has not systematically defined, or proposed comprehensive evaluation frameworks for them. Additionally, no existing work has provided benchmark datasets for evaluating JIR services or developed practical JIR systems for real-world applications.

To fill these gaps, we formalize the JIR task as a partially observable markov decision process, readily transferrable into an AI JIR agent development pipeline (Section \ref{sec:task_formulation}), and propose the evaluation framework for JIR systems (Section \ref{sec:evaluation_metrics}). Moreover, we introduce \jirarena, the first multimodal JIR benchmark with real, diverse, extensive and challenging information-seeking scenarios to assess the \textit{precision}, \textit{recall}, \textit{timeliness} and \textit{relevance} of information recommendations from JIR systems.

The construction of \jirarena involves two primary stages: a user information need simulation stage to generate user queries, and an information retrieval-based JIR instance completion stage to address these queries. This process is challenging, because: \textit{i)} the subjectivity of information need annotations—whether generated by individual annotators or large foundation models—fails to capture the distribution of user information needs, making it unsuitable as a benchmark standard; and \textit{ii)} ensuring reproducibility of evaluation results is difficult due to uncontrollable variables in the JIR instance generation pipeline, such as the ever-growing web-scale document repositories or the evolving capabilities of content-generation models deployed. To address these, we propose: \textit{i)} employing multi-entity, multi-turn simulations to approximate the user information need distribution. Specifically, for a given video, multiple large foundation models and human annotators collaboratively propose information needs to ensure an exhaustive set of needs, with an additional voting step that determines the likelihood of each need being expressed; \textit{ii)} defining the quality of a single JIR instance based on its performance in retrieving information from static knowledge bases to satisfy the information need, thereby eliminating the impact of the varied retrieval repositories or content-generation models of each JIR system to be evaluated; and \textit{iii)} incorporating multi-entity verification at critical steps to ensure the objectivity and generality of the benchmark dataset. \textbf{Overall}, \jirarena encompasses 34 multimedia scenes totaling 831 minutes, focusing on information-intensive scenarios such as lectures and conference talks. Lecture topics span diverse domains, including computer science, neuroscience, finance, mathematics, and chemistry, while conference talks cover subfields of computer science such as AI, programming, cybersecurity, and educational technologies, benefiting from the rapid iteration and accessibility of scholarly content in this field.

Additionally, we develop a prototypical JIR system with streaming inputs as \jirarena's baseline, with the following findings: \textit{i)} JIR systems based on large foundation models can reasonably simulate user information needs (achieving good precision) but often fail to comprehensively cover all user needs (resulting in low recall), especially for information highly likely to be required by most users; and \textit{ii)} effective content retrieval is crucial to satisfying user information needs, yet the commonly used retrieval models underperform in this context, indicating plenty of room for future research.

In summary, our contributions are:

\begin{itemize}
\item We introduce a new promising intelligent information service paradigm:  Just-in-time Information Recommendation (JIR), which emphasizes leveraging AI to minimize user effort.   
    \item We formalize the JIR task and establish an evaluation methodology for JIR with multiple new measures, layout out a theoretical foundation for studying JIR.   
    \item We create \jirarena, the first realistic multimodal benchmark dataset for evaluating JIR systems' utility, featuring diverse, extensive, and representative testing scenarios. 
    \item We implement an extensible baseline JIR system, obtain initial empirical performance results of this prototypical JIR system, and provide insights for future research.
    \item We facilitate future research in this new topic by fully releasing our code and data.
\end{itemize}
\section{JIR Task Formulation}
\label{sec:task_formulation}

Machines discretize the physical world for processing, allowing us to formalize the JIR task as a Partially Observable Markov Decision Process (POMDP) with discrete time framework: $\langle O, S, A, T, R, \Omega, \lambda \rangle$, where the time unit is the machine's discretization scale.

For JIR, an observation $o \in O$ represents the current multimodal context $c \in C$, the observable aspects of the user's evolving persona $u \in U$, and the knowledge bases $K$ to help issue accurate and useful recommendations. Specifically, $c$ includes any visual, auditory, and sensory data consistent with the user's perspective, and $u$ covers the user's \textit{background} $b \in B$ (e.g., experiences, knowledge, traits, etc.) and their \textit{goals} $g \in G$. In other words, the tuple $o=\langle c, \langle b, g\rangle, K\rangle$ forms the typical input of a JIR system. Correspondingly, a state $s \in S$ represents the complete environment, including any world changes, the underlying user persona evolution, and accessible/inaccessible knowledge bases. 

An action $a \in A$ involves initiating/terminating one/more JIR instance(s), or remaining passive when no information needs to be provided. To be specific, each JIR instance $\iota$'s minimal required field is $\iota=\langle t_s, q, Ref, t_e \rangle$, where $t_s$/$t_e$ is the start/end time, $q$ is the user information need in the form of query, and $Ref$ denotes the reference list for making a recommendation to satisfy the need. Formally, the JIR system's output actions include: \textit{i)} $a=\alpha(\iota)$, which creates a $\iota$ with the value $t_s$, $q$, and $Ref$ set, \textit{ii)} $a=\omega(\iota)$, which assigns the $t_e$ to the $\iota$, or iii) $a=\Delta$, which tells the system to stay idle.

The transition function $T : S \times A \rightarrow S$ models state changes based on the current state and JIR system actions. $R : S \times A \rightarrow \mathbb{R}$ is the reward function that determines the utility to the user when the JIR system takes an action in the current state, which is measured by the timeliness and situational relevance of the JIR instances established/canceled by the action. $\Omega$ reflects the initial observation (e.g., information-seeking scenarios, the user profile pool, and knowledge bases) distribution. The discounting factor $\lambda$ is typically 1 or close to it, as most JIRs have immediate, independent impacts. 
 
To solve JIR tasks, a JIR system must design a decision policy $\pi(a_t|\{o_0, o_1, ..., o_t\})$ that maximizes the expected discounted return $\mathbb{E}_{o\sim\Omega}[\sum_{t=0}^{T}\lambda^tR_{t+1}|o_0=o]$, essentially aiming to provide the right information at the right time.

As the first work in the JIR direction, this paper focuses on \textit{i)} defining the JIR POMDP components, \textit{ii)} standardizing evaluation, and \textit{iii)} establishing \jirarena and a baseline JIR system $\pi_{baseline}$.
\section{\jirarena Benchmark Dataset Curation}
\begin{figure*}[t]
    \centering
    \includegraphics[trim={0 0.07cm 0 0}, width=1.0\textwidth]{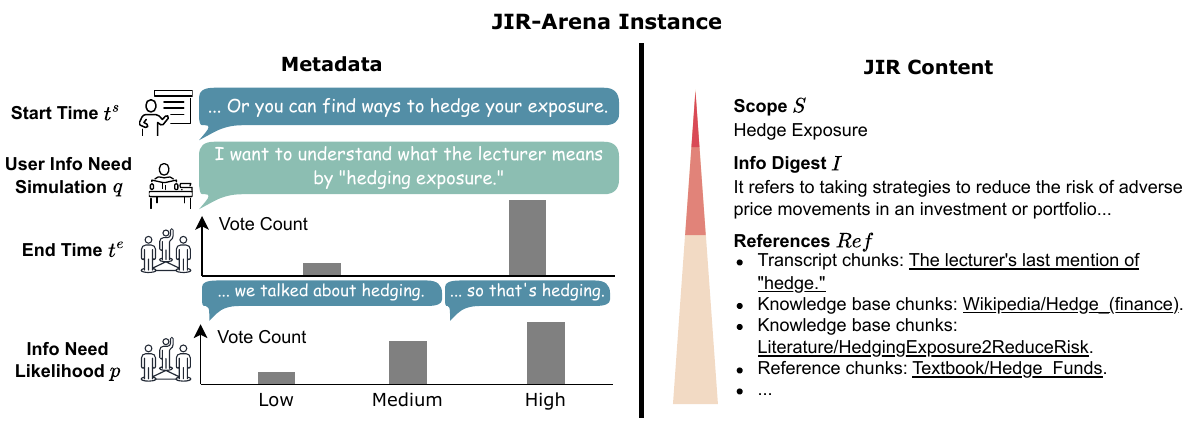}
    \caption{\label{fig:jir_instance}
    A usable form of a JIR instance, with $\langle t_s, q, Ref, t_e \rangle$ to be the minimally required field for measuring the JIR's quality, $p$ for characterizing the need distribution, and $\langle S, I \rangle$ for system display.
    }
\end{figure*}
\begin{figure*}[t]
    \centering
    \includegraphics[trim={0 0.07cm 0 0}, width=1.0\textwidth]{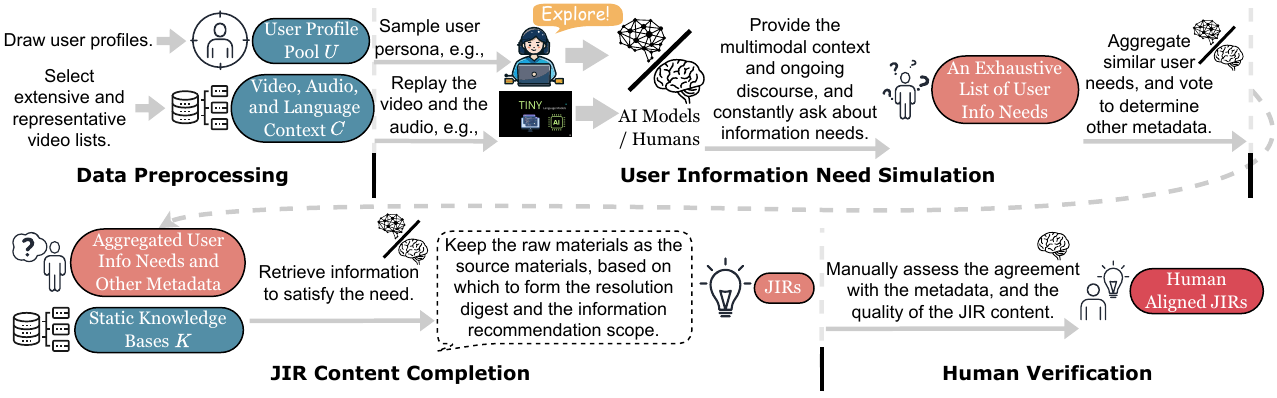}
    \caption{\label{fig:dataset_curation}
    Our pipeline of constructing the \jirarena benchmark dataset.
    }
\end{figure*}
In this section, we introduce the construction methodology of the benchmark dataset. The dataset comprises multiple independent scenes, each containing a collection of JIR instances. As illustrated in Figure \ref{fig:jir_instance}, each JIR instance consists of two components: metadata and JIR content. The metadata component primarily includes the user information need simulation $q$, start time $t_s$ and end time $t_e$. The information need likelihood characterizes $p$ the distribution of simulated user information needs. The JIR content component hierarchically encompasses the scope $S$ and information digest $I$ designed for system display. The references $Ref$ includes a list of source materials that ensures the JIR is formed accurately and faithfully. It is important to note that $t_s$, $q$, $Ref$, and $t_e$ are the minimal required attributes for a JIR system to submit results that can be evaluated in \jirarena. Fields $S$ and $I$ are excluded because their quality are highly dependent on the performance of the generation models used by the JIR system, making controlled variable analysis infeasible. The entire construction process is depicted in Figure \ref{fig:dataset_curation}, and each stage is elaborated in the subsequent sections.

\subsection{Stage 1: Data Preprocessing}
In this stage, we enrich the user profile pool $U$, compile testing scenes $C$, and decide the static knowledge bases $K$. For the user profile pool $U$, human annotators are required to be domain experts likely to have firsthand experience with the tested scenarios in \jirarena. For large foundation models, a role-playing prompt simulates the models as active participants in the given scenarios, instructing the models to request necessary information when needed. Regarding testing scenarios $C$, we focus on two categories characterized by dense information needs: academic lectures and conference talks. Real-world instances of these scenarios are sourced from \textsc{YouTube}, including recorded lectures, talks, and virtual paper explanation videos. The video list is curated by identifying high-subscriber \textsc{YouTube} playlists spanning diverse domains, such as computer science, neuroscience, finance, mathematics, and chemistry for lectures, as well as subtopics within computer science conferences, such as AI and information security. From each playlist, we select the most-viewed videos to ensure scene representativeness. For knowledge bases, we use the Wikipedia dump \citep{petroni-etal-2021-kilt} covering relevant annotations of academic concepts, scholars, and related books, etc., as the basic source. Additionally, for lectures, we add the textbook where applicable. For conference talks, we incorporate arXiv paper titles, abstracts \citep{kaggle_arxiv_dataset}, and, if any, the full text of the papers being discussed during the talks. Metadata for these \textsc{YouTube} videos, textbook, and papers is summarized in Appendix \ref{app:video_textbook_paper_metadata}.
\subsection{Stage 2: User Information Need Simulation}
In this stage, we employ both large foundation models and human annotators to simulate user information needs and estimate the need's distribution. It is important to note that information needs from any single individual are highly subjective and tightly coupled with a user’s persona, including their background and goals. For instance, students from computer science and neuroscience disciplines may require different supplementary information during the same biologically-plausible AI talk. We model user information needs as a time-dependent random variable $Q$, where at each time point $t$, $Q$ has a probability of taking a specific value $q$. The measurement objective of \jirarena, is thus defined as the ability of any JIR system to accurately characterize the probability distribution of $Q$, and to address the queries. To this end, we \textit{i)} curate \textit{exhaustive} user information needs by combining multi-entity simulations while ensuring their reasonableness, deduplicating them based on temporal overlap and semantic similarity, and voting on their likelihood $p=P_{t_s}(Q=q)$ to reflect the probability that $q$ is expressed by all users. The domain completeness of $Q$ ensured by the exhaustive need simulation, and the likelihood estimation derived from multi-entity voting, justify \jirarena as a reasonable and valid benchmark for evaluating JIR system performance.

To be specific, we utilize two types of modules to simulate user information needs: large language models (LLMs) and human annotators. The LLMs include \gpt \citep{openai2024gpt4ocard} and \ds \citep{deepseek2024v3}. For LLM-based simulations, the process involves using \textsc{whisper-large-v3} \citep{radford2022whisper} to transcribe audio content and \textsc{NVILA-8B} \citep{liu2024nvila} to narrate video content, resulting in two input modalities: transcription-only (audio-based) and transcription+narrative (audio+video-based). Each modality is segmented into units of five sentences, with the preceding context summarized as background and the current unit designated as ongoing discourse. Then LLMs are prompted to identify user information needs for the ongoing discourse based on the background. For every scene, this process yields two simulated user information need lists corresponding to the two input modalities for each LLM. Human annotators, on the other hand, are instructed to pause videos whenever they have questions, record the questions alongside timestamps, and provide their own simulation lists for each scene. All user information needs identified within a scene are aggregated, deduplicated based on temporal overlap and semantic similarity (with a similarity threshold of 0.75 as determined by \st \citep{reimers-2019-sentence-bert}, prioritizing human annotations over LLM-generated ones), and finalized. Subsequently, \gpt and \ds are tasked with voting on the likelihood of each element in the exhaustive information need list being raised by human audiences, yielding the likelihood $p$. The LLMs also vote to determine the end time $t_e$ for each JIR instance, given the subjectivity of end-time judgments. All LLM prompts are detailed in Appendix \ref{app:user_information_need_simulation_prompts}.

Notably for human annotators, we conduct an quantitative experiment with four human annotators (the authors of this paper) on nine scenes spanning diverse topics (221 minutes in sum). Analysis reveals \textit{i)} significant variance in user information needs across annotators and \textit{ii)} the ability of LLM simulations to cover these high-variance needs. Consequently, for the remaining videos, exhaustive user information needs are generated solely using LLMs. Detailed analysis are given in Appendix \ref{app:human_annotation_analysis}.
\subsection{Stage 3: JIR Content Completion}
In the JIR instance completion stage, we implement a three-layer information retrieval pipeline comprising a classical information retrieval model, a quality check using LLMs, and a human quality check to ensure that the reference list for each JIR instance is useful for addressing the corresponding information query. The relevance score of a document is determined based on the number of retrieval layers it passes (e.g., documents not retrieved by the information retrieval model are assigned a relevance score of 0, while those passing the human quality check receive the highest score 3).

The classical information retrieval layer narrows the search space in the document repository. We employ the hybrid model from the \textsc{Pyserini} library \citep{Lin_etal_SIGIR2021_Pyserini} for document indexing and retrieval, utilizing \textsc{BM25} ranking \citep{robertson1995okapi} for sparse indexing and \textsc{castorini/tct\_colbert-v2-hnp-msmarco} \citep{lin-etal-2021-batch} as the dense indexing model. Document granularity is designed as: Wikipedia and arXiv datasets are chunked and indexed at the entry level, while textbooks and papers discussed are segmented into their smallest meaningful units (e.g., subsections, figures, or tables) for indexing. At this stage, the retrieval model returns the top 10 most relevant documents from each knowledge base (e.g., Wikipedia, textbooks) for a given query.

The LLM quality check layer leverages multiple LLMs to filter out irrelevant documents through contextual reasoning. Specifically, for each query and the associated documents retrieved, \gpt and \ds are prompted to assess whether a document could provide useful information for answering the query. If deemed relevant, the LLMs extract the helpful portions of the document. The prompts used in this process are detailed in Appendix \ref{app:ir_prompts}.

Finally, the human quality check layer determines the final references. For each query, if all LLMs agree that a document contains information relevant to answering the query, the human annotators are tasked with making a final decision based on the query's context. The results of these three retrieval layers are then integrated to finalize the references and to assign the hierarchical relevance scores.

\subsection{Stage 4: Human Verification}

To verify the quality of the curated JIR instances, we perform a manual verification. The four annotators are tasked with judging the outputs of Stage 2. Note that human verification for Stage 3 has been done with its human quality check layer. For Stage 2, two needs are randomly sampled from each of the videos and assigned to each respective annotator. This results in 272 needs judged. The annotators are instructed to rate the generated \textit{Reason}, \textit{Need}, and \textit{Question} on the following scale:
\begin{enumerate}
    \item[(0)] Disagree: The generated content is implausible;
    \item[(1)] Slightly Disagree: The generated content is somewhat implausible;
    \item[(2)] Neutral: The generated content is possibly plausible;
    \item[(3)] Slightly Agree: The generated content is somewhat plausible;
    \item[(4)] Agree: the generated content is plausible.
\end{enumerate}

The annotators judge the \textit{Need} and \textit{Question} conditioned on the generated \textit{Reason}. The average \textit{Reason} score is 3.26, the average \textit{Need} score is 3.774, and the average \textit{Question} score is 3.801. This indicates that the generated reasons are generally plausible, and that the foundation models are able to generate plausible needs and questions given a reason. The statistics of \jirarena is presented in Table \ref{table:dataset_statistics}.

\section{Evaluation Metrics of JIR Systems}
\label{sec:evaluation_metrics}
\begin{figure*}[t]
    \centering
    \includegraphics[trim={0 0.07cm 0 0}, width=1.0\textwidth]{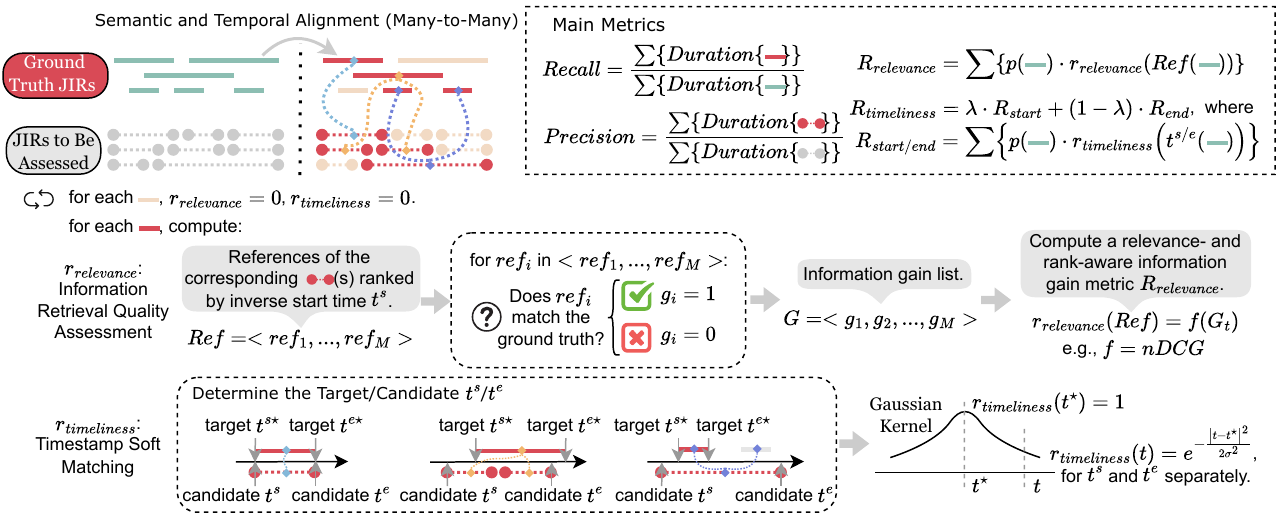}
    \caption{\label{fig:evaluation}
    Evaluation metrics of the JIR systems (top-right): Two global metrics, $Recall$ and $Precision$, based on alignment of ground truth and the candidate answer (top-left), and $R_{relevance}$ and $R_{timeliness}$, aggregated from the relevance and timeliness scores of each JIR instance (bottom).
    }
\end{figure*}
The evaluation of the JIR system considers three main aspects: \textit{i)} accurately inferring user information needs to establish a solid foundation for subsequent retrieval and recommendation tasks; \textit{ii)} avoiding redundant recommendations to minimize the user's cognitive overhead of accessing unnecessary information; and \textit{iii)} ensuring that each JIR instance is contextually relevant, satisfies the information need, and is delivered in a timely manner. Guided by these principles, we propose a set of \textit{scene-wise} metrics to evaluate the JIR systems (Figure \ref{fig:evaluation}). $Recall$ and $Precision$ are designed to address the first two aspects, measuring the coverage of ground truth JIR instances (JIRs) and the effectiveness of JIRs generated by the system, respectively. For the third aspect, we introduce two global metrics, $R_{relevance}$ and $R_{timeliness}$, which aggregate individual scores for relevance and timeliness across ground truth JIRs. The specific computation methods for these metrics are as follows:

\paragraph{$Recall$ and $Precision$:}
To compute $Recall$ and $Precision$, we first perform semantic and temporal alignment between the sets of ground truth JIRs and candidate JIRs. For each ground truth JIR, we identify candidate JIRs with overlapping time intervals and evaluate their semantic similarity. A match is recorded if the semantic content is sufficiently similar (cosine similarity by \st is higher than 0.55), with the matching process flexibly defined as many-to-many to account for the decomposability of user information needs. $Recall$ is then computed as the proportion of ground truth JIRs that are matched, reflecting the system's ability to infer user needs. $Precision$ is calculated as the proportion of candidate JIRs that are matched, assessing the system's efficiency in generating effective JIRs.

\paragraph{$R_{relevance}$:}
This metric evaluates whether the reference lists returned by the JIR system satisfy the user's information need. It aggregates individual relevance scores ($r_{relevance}$) for ground truth JIRs, which are computed using traditional information retrieval metrics. Specifically, for each matched ground truth JIR, the JIR system will provide a ranked list of candidate reference documents. If there are than one matched candidate JIRs, the reference list is ordered by their start times in descending order, as users pay more attention to the newly appeared JIR. Using relevance scores pre-determined in \jirarena, we calculate an information gain list $G$ for the candidate reference list and apply nDCG\footnote{nDCG (normalized Discounted Cumulative Gain) is a metric used to evaluate the quality of ranked results in information retrieval. It measures how well the predicted ranking of search results aligns with the ideal ranking, giving higher importance to relevant items appearing earlier in the list.} \citep{10.1145/582415.582418} to derive $r_{relevance}$. For unmatched ground truth JIRs, $r_{relevance}=0$. The global $R_{relevance}$ is a weighted sum of $r_{relevance}$, where weights are determined by normalizing the likelihood $p$ of each JIR within a given scene.

\paragraph{$R_{timeliness}$:}
This metric assesses the temporal accuracy of the JIR system's recommendations. It balances the alignment of start times ($R_{start}$) and end times ($R_{end}$) through a weighting factor $\lambda$, with each component aggregated across the scene. For individual temporal alignment $R_{timeliness}$, we apply a Gaussian-like kernel to compute a fuzzy match between the ground truth and candidate time points. Specifically, for each matched ground truth JIR, we determine the earliest start time and latest end time among the corresponding candidate JIRs as the timestamp to be evaluated. The timeliness score is then computed as $r_{timeliness} = e^{-\frac{|t - t^\star|^2}{2\sigma^2}}$, where $t^\star$ is the ground truth time, t is the candidate time, and $\sigma$ (set to 3 seconds) represents the tolerance for timing errors. This kernel ensures a smooth decay in score as temporal discrepancies increase. For unmatched ground truth JIRs, $r_{timeliness} = 0$. The global $R_{timeliness}$ is computed as a weighted sum of $r_{timeliness}$, using the same weighting scheme as $R_{relevance}$.
\section{Baseline JIR Systems}

In this section, we describe our baseline JIR system. Our goal with this system was to mimic the real-world instantiation of a JIR system - namely, given streaming multimodal input consistent with the human, can the system predict and resolve the right information needs at the right time?

As as a prototypical JIR system, we focus on the most informative text modality as the input. The full pipeline is illustrated in Figure \ref{fig:baseline}, with two main functioning modules: the \textit{generative model} to predict information needs, and the \textit{retrieval model} to retrieve relevant documents. To be specific, we first segment each transcript into two-minute chunks, and then passed each chunk to state-of-the-art generative models, including \gpt, \ds, \claude \citep{anthropic2024claude3}, and \gemini~\citep{team2023gemini}. Each model has been prompted to generate questions (if any) that a curious listener might have, and then to label these questions with the corresponding start and end timestamps from the transcript. Following the need generation, we then use each resulting questions a query to search over the reference dataset. For the retrieval model, search is conducted using \textsc{OpenSearch}~\citep{opensearch}, where each reference document segment has been pre-processed and indexed. The search formula used is the default instantiation of BM25. We return the top ten best matches to each query. Details and prompts are in Appendix \ref{app:baseline}.

\section{Results}
\begin{table}[t]
\centering
\caption{Baseline JIR systems' performance on \jirarena.}
\renewcommand{\arraystretch}{1.2}
{\scriptsize
\begin{tabular}{l|l|cccccc}
\toprule
\textbf{Generative Model} & \textbf{Retrieval Model} & \textbf{$Recall$} & \textbf{$Precision$} & $R_{relevance}$ & $R_{start}$ & $R_{end}$ & $R_{timeliness}$ \\
\midrule
\gpt & \multirow{4}{*}{\textsc{OpenSearch}} & 0.406 & 0.694 & 0.015 & 0.752 & 0.545 & 0.731 \\
\ds &                      & 0.396 & \textbf{0.716} & 0.014 & 0.737 & 0.548 & 0.719 \\
\claude &                      & 0.327 & 0.696 & \textbf{0.016} & 0.677 & 0.546 & 0.664 \\
\gemini &                      & \textbf{0.429} & 0.694 & 0.014 & \textbf{0.754} & \textbf{0.573} & \textbf{0.736} \\
\bottomrule
\end{tabular}}

\label{table:baseline_performance}
\end{table}

The main results are summarized in Table \ref{table:baseline_performance}, with all metrics averaged over scene durations. For ground truth JIR w/o matched candidates, the $r_{relevance}$ and $r_{timeliness}$ are zero. To prevent $Recall$ from influencing the analysis of relevance of retrieval and timeliness of the system, the $R_{relevance}$ and $R_{timeliness}$ presented are calculated solely for matched ground truth JIRs. Among the evaluated systems, the generative model \gemini achieves the highest $Recall$ (indicating its comprehensive prediction of information needs), and the best \( R_{timeliness} \) (showing reasonable timing in displaying and fading out JIRs). \claude gets the highest \( R_{relevance} \) (indicating the greatest utility of retrieved documents). \ds achieves the best $Precision$ (indicating minimal irrelevant or redundant JIR predictions). However, the overall performance of all JIR systems falls short of practical application standards, underscoring the necessity of using \jirarena to drive further improvements in JIR system development. The analysis follows below.

\paragraph{Can Generative Models Reasonably Infer User Information Needs?}
Error analysis reveals that although the JIR systems exhibit decent performance in predicting user information needs (as reflected in their $Recall$ scores), they often struggle to match needs that are highly likely to be raised by users. For instance, \gemini, which achieves the highest scores overall, shows average $Recall$ values segmented by likelihood \( p \) as follows: 0.37 for \( 0.7 \leq p < 0.8 \), 0.44 for \( 0.8 \leq p < 0.9 \), and 0.48 for \( 0.9 \leq p \leq 1.0 \). The relatively consistent performance across likelihood levels suggests that the baseline JIR systems’ user simulation approaches remain underdeveloped. Potential improvement directions include adding multi-modal inputs and reducing content window sizes for more fine-grained predictions (though with increased computational costs). Additionally, in the future, to achieve greater personalization, JIR systems could benefit from the collection of more comprehensive user profiles.

\paragraph{Can Retrieval Models Retrieve Relevant Documents to Address Information Needs?}

The retrieval model primarily influences \( R_{relevance} \), and we observe that despite employing widely-used retrieval systems, their ability to retrieve documents relevant to user information needs is poor. Qualitative analysis reveales that this limitation is largely due to the highly context-dependent nature of user queries. For example, in a scene where a lecturer states that a certain approach is ``not the way forward,'' the corresponding information need (passed to the retrieval model as a query) might be, ``Which approach is not the way forward, and why is it unsuitable?'' When only the information need is used as the query, w/o contextual information, it often appears vague, leading to low retrieval quality. We recommend that future JIR system development incorporate the contextual background of information needs during retrieval, thereby enabling \textit{contextualized} information retrieval.

\paragraph{Are JIR System Recommendations Timely?}

We find that the average timing error of the systems is approximately 10 seconds (about two sentences' delay after the content triggering the information need appears in the environment). When the information need is accurately predicted, this delay reduces to approximately 5 seconds (about one sentence). Although these delays exist, they fall within the range of human-perceived acceptable error. To further improve timeliness, we recommend enhancing the accuracy of user information need simulation and processing input at finer time granularity within the JIR system.
\section{Related Work}
\label{app:related_work}

The closest work is \citep{rhodes2000just}, which introduced the concept of Just-In-Time Information Retrieval (JITIR) Agents. However, JITIR
has not attracted much attention since it was proposed more than two decades ago, likely due to the lack of a benchmark for evaluating algorithms. The JIR vision we proposed is closely related to JITIR but with three distinctions: \textit{i)} We emphasize recommendation (going beyond retrieval), aiming to maximize the intelligence of a JIR system and create more room for AI research (e.g., developing algorithms for detecting a user's need in real time).
However, JIR would also allow a user to provide hints to facilitate recommendation.  \textit{ii)} The JIR task formulation includes predicting the user need time window, which was not included in JITIR. \textit{iii)} While JITIR is a vague concept, we mathematically define JIR as a Partially Observable Markov Decision Process (POMDP). As JIR is more general than JITIR, the \jirarena benchmark we created can also facilitate research on JITIR. JIR is also related to a few other tasks proposed in the existing work, including background linking~\citep{soboroff2020trec}, document-to-slide generation\citep{fu2022doc2ppt}, gaze and vocal interaction
\citep{khan2022integrating}, 
SearchBot~\citep{andolina2018searchbot}, and visual caption
~\citep{liu2023visual}.
The traditional measures used for search and recommendation evaluation~\citep{sanderson2010test} are inadequate for evaluating the new JIR task, and we propose a new evaluation methodology and introduce multiple new measures for evaluating JIR. 
Many datasets were created for evaluating both search and recommendation tasks, mostly via initiatives such as TREC~\citep{voorhees2005trec}, CLEF~\citep{ferro2015clef}, and NTCIR~\citep{oard2019celebrating}, open datasets such as those for evaluating recommendation algorithms in domains such as movies~\citep{harper2015movielens}, music~\citep{schedl2016lfm}, and products~\citep{jin2023amazon}. Our new dataset differs from all the previous datasets in that it includes a multi-modality conversational context as a basis for recommendation. 
Some work created datasets more closely related to our dataset in annotation of questions, including   proactive search \citep{samarinas2024procis}, question generation based on lectures~\citep{chen2018learningq}, TED talks~\citep{westera2020ted}, and news articles~\citep{ko2020inquisitive}. Compared with them, our dataset is much more comprehensive including additional information such as time window and relevant source references. LLMs are often exploited as judge for automating data annotations~\citep{gu2024survey}, including making relevance judgments for evaluating search algorithms~\citep{faggioli2023perspectives} and simulating users for evaluating conversational assistants~\citep{INR-098}. We also leverage LLMs but for creating annotations for a new task that has not been evaluated before.

\section{Discussions}
As AI continues to advance in sophistication and its deployment costs decrease, the prospect of individuals having access to customized AI assistants is becoming increasingly tangible in the near future. This shift holds profound implications, as it enables the democratization of services that were once prohibitively expensive, highly specialized, and exclusively personalized. Among such services, intelligent and tailored information recommendation systems, exemplified by the JIR systems, serve as a compelling example. In this section, we discuss potential future directions for the JIR systems, reflecting three general challenges that have broader relevance to the emerging class of \textit{customized intelligent services} enabled by AI. 

\paragraph{Contextualization}
Contextualization refers to the ability of the JIR system to provide recommendations that fully account for the user’s situational context, including their physical environment and even physiological cues. For instance, subtle micro-behaviors such as a furrowed brow indicating confusion or prolonged focus on a specific formula may help the system better infer the user’s information needs. We anticipate that contextualization will be a foundational capability of future JIR systems, and AI service systems in general. However, its development is currently constrained by several technological bottlenecks in AI research. These include embodied intelligence (AI’s limited capacity to understand and interact with the evolving physical environment), multimodal content understanding and generation (difficulties in processing sensory inputs like vision, sound, and smell that align with or surpass human perception), and context window limitations (the absence of robust memory mechanisms to manage working memory and episodic memory effectively), etc. Advances in any of these critical areas are likely to enhance the capabilities of JIR systems.

\paragraph{Personalization}
Personalization in JIR systems involves tailoring recommendations to the unique background, goals, and thus diverse needs of individual users. The primary challenge of AI assistant personalization lies in user simulation~\citep{INR-098}, which requires the AI agent to not only analyze observable data, such as the user's browsing or travel histories, but also infer their latent, unarticulated needs that objectively exist, which can be exemplified by the critical role of the user information need simulation in the JIR system development. We argue that high recall is the most crucial requirement for effective user need simulation, ensuring that the system considers and covers a comprehensive range of all potential user demands. The second important aspect is achieving high precision in recommendations through contextualized inference, as irrelevant or redundant suggestions can detract from the user's experience. In action, balancing comprehensive need identification with selective, high-precision recommendations will be a defining feature of effective personalized JIR systems.

\paragraph{Proactive Service}  
The concept of proactive service emphasizes the \textit{proactive} nature of the JIR system, distinguishing it from mainstream AI agents that typically respond to user demands only when explicitly requested \citep{openai2024gpt4ocard,deepseek2024v3,anthropic2024claude3,team2023gemini}. Instead, the JIR system integrates services into daily interactions in the manner of an everyday assistant or companion. For instance, imagine a scenario where, while having breakfast, your smart glasses proactively inquire whether you would like an update on the previous day's AI technology advancements and stock prices of leading tech companies. To clarify, the proactive services advocated by JIR emphasize minimizing a user's effort, but a JIR system can also accommodate that a user provides additional input such as a query to further clarify the need. Although these systems remain tools within the user's life, their inputs are far richer, encompassing the user's full experiences; and their design objective shifts from merely addressing user-presented needs to anticipating and preemptively fulfilling them in a non-intrusive, non-disruptive manner. Achieving this, however, necessitates that the AI assistant remain constantly online to absorb user information. As a result, the trustworthiness of such systems, particularly in terms of safety, fairness, and privacy, becomes a critical concern in developing this form of intelligence. From the perspective of optimizing AI-human collaboration, there is generally a tradeoff between minimization of a user's effort and trustworthiness of the AI system. As the AI systems are designed to take over more work from the user to reduce their needed effort, there will be an increased concern and risk of reliability on the AI system.  
\section{Conclusions}
This work is the first to introduce Just-in-time Information Recommendation (JIR) to the broader research community, highlighting its potential as a transformative information service. JIR focuses on recommending relevant information to users at the most opportune moments, in a minimally intrusive manner, to address their information gaps and support decision-making. With technological advancements increasingly enabling the feasibility of JIR systems, this helpful service is poised to reshape how individuals interact with information, unlocking commercial opportunities and driving new research directions.

Our contributions include a formal mathematical definition of the JIR task, the establishment of a comprehensive set of evaluation metrics tailored to the expectations of well-functioning JIR systems, and the creation of \jirarena, the first multimodal JIR benchmark dataset, which encompasses diverse, representative, and challenging scenes necessitating JIR services. Additionally, we develop a prototype JIR system and conduct performance evaluations and error analyses on \jirarena to propose avenues for future system improvements. To facilitate further research in this promising domain, we have made all associated code and data fully accessible. Moving forward, we anticipate increased exploration of JIR application scenes and advancements in developing state-of-the-art JIR systems regarding the basic performance and JIR personalization.

\bibliography{references}
\bibliographystyle{plainnat}


\appendix
\section{Metadata of Pre-processed Videos, Textbook and Papers}
\label{app:video_textbook_paper_metadata}
\subsection{Video Metadata}
\begin{itemize}
\item \textbf{Title}: \href{https://www.youtube.com/watch?v=IFKnq9QM6_A}{Elements and atoms | Atoms, compounds, and ions | Chemistry | Khan Academy}\\
\textbf{Duration}: 13m 9s\\
\textbf{Channel}: \href{https://www.youtube.com/channel/UC4a-Gbdw7vOaccHmFo40b9g}{Khan Academy}\\
\textbf{Views}: 4,192,084\\
\textbf{YouTube ID}: IFKnq9QM6\_A

\item \textbf{Title}: \href{https://www.youtube.com/watch?v=Jsiy4TxgIME}{Basic trigonometry | Basic trigonometry | Trigonometry | Khan Academy}\\
\textbf{Duration}: 9m 17s\\
\textbf{Channel}: \href{https://www.youtube.com/channel/UC4a-Gbdw7vOaccHmFo40b9g}{Khan Academy}\\
\textbf{Views}: 4,131,663\\
\textbf{YouTube ID}: Jsiy4TxgIME

\item \textbf{Title}: \href{https://www.youtube.com/watch?v=2f7YwCtHcgk}{Introduction to cellular respiration | Cellular respiration | Biology | Khan Academy}\\
\textbf{Duration}: 14m 19s\\
\textbf{Channel}: \href{https://www.youtube.com/channel/UC4a-Gbdw7vOaccHmFo40b9g}{Khan Academy}\\
\textbf{Views}: 3,249,406\\
\textbf{YouTube ID}: 2f7YwCtHcgk

\item \textbf{Title}: \href{https://www.youtube.com/watch?v=MXJ-zpJeY3E}{The World's Best Mathematician (*) - Numberphile}\\
\textbf{Duration}: 10m 57s\\
\textbf{Channel}: \href{https://www.youtube.com/channel/UCoxcjq-8xIDTYp3uz647V5A}{Numberphile}\\
\textbf{Views}: 7,575,385\\
\textbf{YouTube ID}: MXJ-zpJeY3E

\item \textbf{Title}: \href{https://www.youtube.com/watch?v=G2_Q9FoD-oQ}{158,962,555,217,826,360,000 (Enigma Machine) - Numberphile}\\
\textbf{Duration}: 11m 51s\\
\textbf{Channel}: \href{https://www.youtube.com/channel/UCoxcjq-8xIDTYp3uz647V5A}{Numberphile}\\
\textbf{Views}: 6,318,079\\
\textbf{YouTube ID}: G2\_Q9FoD-oQ

\item \textbf{Title}: \href{https://www.youtube.com/watch?v=BRRolKTlF6Q}{Problems with Zero - Numberphile}\\
\textbf{Duration}: 13m 0s\\
\textbf{Channel}: \href{https://www.youtube.com/channel/UCoxcjq-8xIDTYp3uz647V5A}{Numberphile}\\
\textbf{Views}: 5,758,032\\
\textbf{YouTube ID}: BRRolKTlF6Q

\item \textbf{Title}: \href{https://www.youtube.com/watch?v=2s4TqVAbfz4}{Perfect Shapes in Higher Dimensions - Numberphile}\\
\textbf{Duration}: 26m 19s\\
\textbf{Channel}: \href{https://www.youtube.com/channel/UCoxcjq-8xIDTYp3uz647V5A}{Numberphile}\\
\textbf{Views}: 5,326,862\\
\textbf{YouTube ID}: 2s4TqVAbfz4

\item \textbf{Title}: \href{https://www.youtube.com/watch?v=AuA2EAgAegE}{e (Euler's Number) - Numberphile}\\
\textbf{Duration}: 10m 42s\\
\textbf{Channel}: \href{https://www.youtube.com/channel/UCoxcjq-8xIDTYp3uz647V5A}{Numberphile}\\
\textbf{Views}: 4,811,422\\
\textbf{YouTube ID}: AuA2EAgAegE

\item \textbf{Title}: \href{https://www.youtube.com/watch?v=OkmNXy7er84}{The hardest problem on the hardest test}\\
\textbf{Duration}: 11m 15s\\
\textbf{Channel}: \href{https://www.youtube.com/channel/UCYO_jab_esuFRV4b17AJtAw}{3Blue1Brown}\\
\textbf{Views}: 15,931,563\\
\textbf{YouTube ID}: OkmNXy7er84

\item \textbf{Title}: \href{https://www.youtube.com/watch?v=v68zYyaEmEA}{Solving Wordle using information theory}\\
\textbf{Duration}: 30m 38s\\
\textbf{Channel}: \href{https://www.youtube.com/channel/UCYO_jab_esuFRV4b17AJtAw}{3Blue1Brown}\\
\textbf{Views}: 10,981,963\\
\textbf{YouTube ID}: v68zYyaEmEA

\item \textbf{Title}: \href{https://www.youtube.com/watch?v=GNcFjFmqEc8}{But why is a sphere's surface area four times its shadow?}\\
\textbf{Duration}: 15m 50s\\
\textbf{Channel}: \href{https://www.youtube.com/channel/UCYO_jab_esuFRV4b17AJtAw}{3Blue1Brown}\\
\textbf{Views}: 8,278,649\\
\textbf{YouTube ID}: GNcFjFmqEc8

\item \textbf{Title}: \href{https://www.youtube.com/watch?v=IHZwWFHWa-w}{Gradient descent, how neural networks learn | DL2}\\
\textbf{Duration}: 20m 33s\\
\textbf{Channel}: \href{https://www.youtube.com/channel/UCYO_jab_esuFRV4b17AJtAw}{3Blue1Brown}\\
\textbf{Views}: 7,732,696\\
\textbf{YouTube ID}: IHZwWFHWa-w

\item \textbf{Title}: \href{https://www.youtube.com/watch?v=wvXDB9dMdEo}{1. Introduction, Financial Terms and Concepts}\\
\textbf{Duration}: 60m 30s\\
\textbf{Channel}: \href{https://www.youtube.com/channel/UCEBb1b_L6zDS3xTUrIALZOw}{MIT OpenCourseWare}\\
\textbf{Views}: 7,584,607\\
\textbf{YouTube ID}: wvXDB9dMdEo

\item \textbf{Title}: \href{https://www.youtube.com/watch?v=HtSuA80QTyo}{Lecture 1: Algorithmic Thinking, Peak Finding}\\
\textbf{Duration}: 53m 21s\\
\textbf{Channel}: \href{https://www.youtube.com/channel/UCEBb1b_L6zDS3xTUrIALZOw}{MIT OpenCourseWare}\\
\textbf{Views}: 5,733,981\\
\textbf{YouTube ID}: HtSuA80QTyo

\item \textbf{Title}: \href{https://www.youtube.com/watch?v=ghZRzOb_bZo}{Lecture 25 — Probabilistic Topic Models  Expectation Maximization Algorithm - Part 3 | UIUC}\\
\textbf{Duration}: 6m 25s\\
\textbf{Channel}: \href{https://www.youtube.com/channel/UC5zx8Owijmv-bbhAK6Z9apg}{Artificial Intelligence - All in One}\\
\textbf{Views}: 1,574\\
\textbf{YouTube ID}: ghZRzOb\_bZo

\item \textbf{Title}: \href{https://www.youtube.com/watch?v=SoZStBaLbws}{Lecture 42 — Text Categorization  Evaluation - Part 2 | UIUC}\\
\textbf{Duration}: 10m 52s\\
\textbf{Channel}: \href{https://www.youtube.com/channel/UC5zx8Owijmv-bbhAK6Z9apg}{Artificial Intelligence - All in One}\\
\textbf{Views}: 520\\
\textbf{YouTube ID}: SoZStBaLbws

\item \textbf{Title}: \href{https://www.youtube.com/watch?v=Hpw31YTq_yI}{NeurIPS 2018: Keynote on What Bodies Think About}\\
\textbf{Duration}: 56m 51s\\
\textbf{Channel}: \href{https://www.youtube.com/channel/UCvqEpkx-HQ2nDMT-ob-AADg}{ICML IJCAI ECAI 2018 Conference Videos}\\
\textbf{Views}: 1,600\\
\textbf{YouTube ID}: Hpw31YTq\_yI

\item \textbf{Title}: \href{https://www.youtube.com/watch?v=AjNkHz3fRjs}{Are Time Series Foundation Models Ready to Revolutionize Predictive Building Analytics?}\\
\textbf{Duration}: 15m 44s\\
\textbf{Channel}: \href{https://www.youtube.com/channel/UCPyA0XmU6aS4JCwVoIBTmIQ}{Association for Computing Machinery (ACM)}\\
\textbf{Views}: 276\\
\textbf{YouTube ID}: AjNkHz3fRjs

\item \textbf{Title}: \href{https://www.youtube.com/watch?v=0UGhxVwRmcU}{WebSci'24: Keynote by Dirk Hovy}\\
\textbf{Duration}: 60m 16s\\
\textbf{Channel}: \href{https://www.youtube.com/channel/UCPyA0XmU6aS4JCwVoIBTmIQ}{Association for Computing Machinery (ACM)}\\
\textbf{Views}: 162\\
\textbf{YouTube ID}: 0UGhxVwRmcU

\item \textbf{Title}: \href{https://www.youtube.com/watch?v=TZhUWPzAvJc}{IOHanalyzer: Detailed Performance Analyses for Iterative Optimization Heuristics}\\
\textbf{Duration}: 7m 3s\\
\textbf{Channel}: \href{https://www.youtube.com/channel/UCPyA0XmU6aS4JCwVoIBTmIQ}{Association for Computing Machinery (ACM)}\\
\textbf{Views}: 268\\
\textbf{YouTube ID}: TZhUWPzAvJc

\item \textbf{Title}: \href{https://www.youtube.com/watch?v=ThXTZQN_m3E}{EAAMO'22: Liquid Democracy in Practice: An Empirical Analysis of its Epistemic Performance}\\
\textbf{Duration}: 19m 28s\\
\textbf{Channel}: \href{https://www.youtube.com/channel/UCPyA0XmU6aS4JCwVoIBTmIQ}{Association for Computing Machinery (ACM)}\\
\textbf{Views}: 263\\
\textbf{YouTube ID}: ThXTZQN\_m3E

\item \textbf{Title}: \href{https://www.youtube.com/watch?v=k_gGx9ObXeM}{First Talk IDC 2022 Europe Best Paper Video}\\
\textbf{Duration}: 14m 24s\\
\textbf{Channel}: \href{https://www.youtube.com/channel/UCPyA0XmU6aS4JCwVoIBTmIQ}{Association for Computing Machinery (ACM)}\\
\textbf{Views}: 233\\
\textbf{YouTube ID}: k\_gGx9ObXeM

\item \textbf{Title}: \href{https://www.youtube.com/watch?v=1CJvt4OjWgs}{Document Clustering vs Topic Models: A Case Study}\\
\textbf{Duration}: 16m 31s\\
\textbf{Channel}: \href{https://www.youtube.com/channel/UCPyA0XmU6aS4JCwVoIBTmIQ}{Association for Computing Machinery (ACM)}\\
\textbf{Views}: 1,300\\
\textbf{YouTube ID}: 1CJvt4OjWgs

\item \textbf{Title}: \href{https://www.youtube.com/watch?v=Kkx--T5NUy4}{KDD Keynote Talk--On the Nature of Data--Jeffrey D Ullman}\\
\textbf{Duration}: 60m 13s\\
\textbf{Channel}: \href{https://www.youtube.com/channel/UCPyA0XmU6aS4JCwVoIBTmIQ}{Association for Computing Machinery (ACM)}\\
\textbf{Views}: 2,200\\
\textbf{YouTube ID}: Kkx--T5NUy4

\item \textbf{Title}: \href{https://www.youtube.com/watch?v=Dxit4BOO0y8}{My Mouse, My Rules: Privacy Issues of Behavioral User Profiling via Mouse Tracking}\\
\textbf{Duration}: 14m 52s\\
\textbf{Channel}: \href{https://www.youtube.com/channel/UCPyA0XmU6aS4JCwVoIBTmIQ}{Association for Computing Machinery (ACM)}\\
\textbf{Views}: 690\\
\textbf{YouTube ID}: Dxit4BOO0y8

\item \textbf{Title}: \href{https://www.youtube.com/watch?v=7XH53R675Qw}{Simulation of cognitive agents to explore occupants’ wayfinding experience in future buildings}\\
\textbf{Duration}: 27m 48s\\
\textbf{Channel}: \href{https://www.youtube.com/channel/UCPyA0XmU6aS4JCwVoIBTmIQ}{Association for Computing Machinery (ACM)}\\
\textbf{Views}: 486\\
\textbf{YouTube ID}: 7XH53R675Qw

\item \textbf{Title}: \href{https://www.youtube.com/watch?v=27zuReojDVw}{SIGCOMM 2020 Keynote: Amin Vadhat: Coming of Age in the Fifth Epoch of Distributed Computing}\\
\textbf{Duration}: 49m 46s\\
\textbf{Channel}: \href{https://www.youtube.com/channel/UCPyA0XmU6aS4JCwVoIBTmIQ}{Association for Computing Machinery (ACM)}\\
\textbf{Views}: 3,200\\
\textbf{YouTube ID}: 27zuReojDVw

\item \textbf{Title}: \href{https://www.youtube.com/watch?v=LvQWRp1lwzY}{ACM ICN 2020 - Discovering in-network Caching Policies in NDN Networks from a Measurement Perspect.}\\
\textbf{Duration}: 21m 2s\\
\textbf{Channel}: \href{https://www.youtube.com/channel/UCPyA0XmU6aS4JCwVoIBTmIQ}{Association for Computing Machinery (ACM)}\\
\textbf{Views}: 852\\
\textbf{YouTube ID}: LvQWRp1lwzY

\item \textbf{Title}: \href{https://www.youtube.com/watch?v=9Rxb2px3QcI}{L@S 2020: "Human Languages in Source Code: Auto-Translation for Localized Instruction"}\\
\textbf{Duration}: 11m 43s\\
\textbf{Channel}: \href{https://www.youtube.com/channel/UCPyA0XmU6aS4JCwVoIBTmIQ}{Association for Computing Machinery (ACM)}\\
\textbf{Views}: 354\\
\textbf{YouTube ID}: 9Rxb2px3QcI

\item \textbf{Title}: \href{https://www.youtube.com/watch?v=-dVb6j_Ooe4}{Session 1A - Bipartite TSP in O(1.9999) Time, Assuming Quadratic Time Matrix Multiplication}\\
\textbf{Duration}: 25m 0s\\
\textbf{Channel}: \href{https://www.youtube.com/channel/UCPyA0XmU6aS4JCwVoIBTmIQ}{Association for Computing Machinery (ACM)}\\
\textbf{Views}: 4,500\\
\textbf{YouTube ID}: -dVb6j\_Ooe4

\item \textbf{Title}: \href{https://www.youtube.com/watch?v=HEqQ2_1XRTs}{"Reinforcement Learning for Recommender Systems: A Case Study on Youtube," by Minmin Chen}\\
\textbf{Duration}: 33m 17s\\
\textbf{Channel}: \href{https://www.youtube.com/channel/UCPyA0XmU6aS4JCwVoIBTmIQ}{Association for Computing Machinery (ACM)}\\
\textbf{Views}: 24,000\\
\textbf{YouTube ID}: HEqQ2\_1XRTs

\item \textbf{Title}: \href{https://www.youtube.com/watch?v=J1tx4OZ_wMc}{Phishing Attacks on Modern Android}\\
\textbf{Duration}: 19m 54s\\
\textbf{Channel}: \href{https://www.youtube.com/channel/UCPyA0XmU6aS4JCwVoIBTmIQ}{Association for Computing Machinery (ACM)}\\
\textbf{Views}: 9,100\\
\textbf{YouTube ID}: J1tx4OZ\_wMc

\item \textbf{Title}: \href{https://www.youtube.com/watch?v=TU19Orwu4jE}{TINY LM Agents on Edge Devices: Can We Scale?}\\
\textbf{Duration}: 28m 8s\\
\textbf{Channel}: \href{https://www.youtube.com/channel/UCfOvNb3xj28SNqPQ_JIbumg}{Discover AI}\\
\textbf{Views}: 2,964\\
\textbf{YouTube ID}: TU19Orwu4jE

\item \textbf{Title}: \href{https://www.youtube.com/watch?v=rqXZDJiKyA8}{Large-Scale Student Data Reveal Sociodemographic Gaps in Procrastination Behavior}\\
\textbf{Duration}: 29m 51s\\
\textbf{Channel}: \href{https://www.youtube.com/channel/UCPyA0XmU6aS4JCwVoIBTmIQ}{Association for Computing Machinery (ACM)}\\
\textbf{Views}: 131\\
\textbf{YouTube ID}: rqXZDJiKyA8

\end{itemize}

\subsection{Textbook}
We used the Latex source of the textbook \textit{Text Data Management and Analysis: A Practical Introduction to Information Retrieval and Text Mining} \citep{10.1145/2915031} as the collection of source information related to some lecture videos in \jirarena.

\subsection{Papers}
The papers \jirarena covers include: \cite{lee2018answererquestionersmindinformation}, \cite{lindenbaum2018geometrybaseddatageneration}, \cite{arik2018neuralvoicecloningsamples}, \cite{chan2018likelihoodfreeinferenceframeworkpopulation}, \cite{tatbul2019precisionrecalltimeseries}, \cite{zhang2018generalizingtreeprobabilityestimation}, \cite{chen2019learningoptimizetensorprograms}, \cite{whittington2018generalisationstructuralknowledgehippocampalentorhinal}, \cite{tobar2019bayesiannonparametricspectralestimation}, \cite{ke2018sparseattentivebacktrackingtemporal}, \cite{yi2019neuralsymbolicvqadisentanglingreasoning}, \cite{li2020learningtemporalpointprocesses}, \cite{shivkumar2018probabilisticpopulationcodebased}, \cite{NEURIPS2018_e02af582}, \cite{cen2024transparencyaccountabilitybackdiscussion}, \cite{10.1145/3671127.3698177}, \cite{10.1145/3632775.3661957}, \cite{wang2022iohanalyzerdetailedperformanceanalyses}, \cite{Revel2022LiquidDI}, \cite{10.1145/3501712.3529716}, \cite{10.1145/3491140.3528279}, \cite{10.1145/3501710.3519518}, \cite{10.1145/3498366.3505816}, \cite{10.1145/3503516.3503527}, \cite{Leiva_2021}, \cite{Perera_2020}, \cite{10.1145/3405656.3418711}, \cite{piech2019humanlanguagessourcecode}, \cite{10.1145/3357713.3384264}, \cite{10.1145/3243734.3243778}, \cite{10.1145/3208976.3209026}, \cite{10.1145/3170427.3188403}, \cite{yang2024tinyhelenscurriculumtrainingevaluating}, and \cite{10.1145/3491140.3528285}.

\subsection{\jirarena Dataset Statistics}
\begin{table}[ht]
\centering
\caption{Statistics of different video categories.}
\renewcommand{\arraystretch}{1.2}
{\scriptsize
\begin{tabular}{l|c|c|c|c}
\toprule
\textbf{Category} & \textbf{Scene Num} & \textbf{Avg View Count / Scene} & \textbf{Avg Duration / Scene (min)} & \textbf{Avg JIR Num / Min} \\
\midrule
Lectures          & 16                 & 6,100,530                       & 19.9                                & 13.1                       \\
Conferences       & 18                 & 2,921                           & 28.4                                & 10.4                       \\
\bottomrule
\end{tabular}
}
\label{table:dataset_statistics}
\end{table}

\jirarena includes 2 categories of scenes, totaling 34 of them and spanning 831 mins. We select the most viewed videos (by \textbf{Avg View Count / Scene}) with moderate duration (by \textbf{Avg Duration / Scene (min)}) and of intensive information needs (by \textbf{Avg JIR Num / Min}).
\section{LLM Prompts for User Information Need Simulation}

\label{app:user_information_need_simulation_prompts}

\begin{prompt}{User Information Need Simulation from Transcript}
Analyze this transcript segment for information needs. I will provide you with a summary of the presentation so far and a transcript segment from the presentation. \\

Categorize each information need into one of the following categories: \\

1. Visual References (graphs, images, diagrams, etc.)  \\
2. Technical Terms (jargon, acronyms, formulas, definitions)  \\
3. Data \& Sources (uncited stats, vague claims like "studies show...") \\ 
4. Processes/Methods (unexplained workflows/algorithms) \\ 
5. External Content (papers, tools, historical references without context) \\  
6. Ambiguous Language (vague terms like "many" or "significant")  \\
7. Missing Context (assumed prior knowledge, undefined goals)  \\
8. Instructions/Actions (unclear steps, implied tasks)  \\
9. Code/Formulas (unexplained pseudocode/equations)  \\
10. Future Work (vague next steps, unresolved questions) \\ 
11. Conceptual Understanding (concepts, ideas)\\

I need an exhaustinve list of information needs. Each sentence can need multiple information needs.\\
Return a JSON list with `type`, `subtype`, and `reason` for each need. Ensure the output is strictly JSON-compliant \\

Summary of Presentation So Far:\\
\{\{Background\_Context\}\}\\

Presentation Transcript Segment:\\
\{\{Transcript\_Chunk\}\}\\

Output Format (JSON):\\
\{
  "information\_needs": [
    \{
      "sentence\_id": "sentence\_id",
      "type": "need type",
      "subtype": "need subtype",
      "reason": "reason for need"
    \},
    \{
      "sentence\_id": "sentence\_id",
      "type": "need type",
      "subtype": "need subtype",
      "reason": "reason for need"
    \},
    \{
      "sentence\_id": "sentence\_id",
      "type": "need type",
      "subtype": "need subtype",
      "reason": "reason for need"
    \}
  ]
\}
\end{prompt}

\begin{prompt}{User Information Need Simulation from Transcript and Narrative}
Analyze this description of a presentation for information needs. I will provide you with a summary of the presentation so far and a segment from the presentation. \\

Categorize each information need into one of the following categories: \\

1. Visual References (graphs, images, diagrams, etc.) \\ 
2. Technical Terms (jargon, acronyms, formulas, definitions)  \\
3. Data \& Sources (uncited stats, vague claims like "studies show...")  \\
4. Processes/Methods (unexplained workflows/algorithms)  \\
5. External Content (papers, tools, historical references without context)  \\
6. Ambiguous Language (vague terms like "many" or "significant")  \\
7. Missing Context (assumed prior knowledge, undefined goals)  \\
8. Instructions/Actions (unclear steps, implied tasks)  \\
9. Code/Formulas (unexplained pseudocode/equations)  \\
10. Future Work (vague next steps, unresolved questions) \\ 
11. Conceptual Understanding (concepts, ideas)\\

I need an exhaustinve list of information needs. Each sentence can need multiple information needs.\\
Return a JSON list with `type`, `subtype`, and `reason` for each need.  Ensure the output is strictly JSON-compliant\\

Summary of Presentation So Far:\\
\{\{Background\_Context\}\}\\

Presentation Description Segment:\\
\{\{Presentation\_Chunk\}\}\\

Output Format (JSON):\\
\{
  "information\_needs": [
    \{
      "sentence\_id": sentence\_id
      "type": "need type",
      "subtype": "need subtype",
      "reason": "reason for need"
    \},
    \{
      "sentence\_id": sentence\_id
      "type": "need type",
      "subtype": "need subtype",
      "reason": "reason for need"
    \},
    \{
      "sentence\_id": sentence\_id
      "type": "need type",
      "subtype": "need subtype",
      "reason": "reason for need"
    \},
  ]
\}
\end{prompt}

\begin{prompt}{Summary}
You are an AI assistant repsonsible for generating a detailed summary of all the sentences from a transcript provided to You\\
Ensure the output is strictly JSON-compliant \\

Input:\\

Transcript Sentences:\\
\{\{Sentences\}\}\\

Output Format (JSON):\\
\{
    "summary": "This is the summary of all the sentences",
\}

Explanation of Each Output Field:\\

summary: Detailed Summary of all the sentences as a single paragraph\\
\end{prompt}

\begin{prompt}{Likelihood Judgement}
You are an AI Judge, evaluating how relevant an informational need is to a presentation — as if you were a thoughtful human attending the talk.\\
Your job is to score the need's relevance on a scale from 0 to 10 based on how likely it is that a curious, context-aware human would naturally have this question or need at this exact point in the presentation.\\
The relvance score must be in the range 0 to 10 where:\\

0 – Completely irrelevant: No connection whatsoever to the content.\\
1 – Barely related: Random term overlap; totally misplaced in context.\\
2 – Weakly related: Vague thematic connection, but wouldn't arise from this presentation.\\
3 – Marginally related: A human could get here with effort, but it feels out of place.\\
4 – Somewhat related: On-topic, but not something a typical attendee would care about now.\\
5 – Mildly relevant: Plausible side question, but still feels like a stretch or a detour.\\
6 – Reasonably relevant: A thoughtful listener might ask this, though it's not the most pressing or natural next step.\\
7 – Clearly relevant: A typical, attentive participant could raise this with no prompting. Fits the flow.\\
8 – Strongly relevant: Feels like a helpful and likely next question from a human audience member. Supports or extends what's being discussed.\\
9 – Very relevant: Almost anticipates what the speaker might say next. Shows deep understanding and interest.\\
10 – Perfectly aligned: A human would almost certainly ask this next. Feels like the natural continuation of the discussion.\\

Evaluation Guidelines\\
Imagine yourself as a human audience member who has been following the presentation closely.\\
Consider flow, timing, speaker’s tone, and logical build-up.\\
Use 7 or higher ONLY for needs that a genuinely attentive human would likely raise unprompted.\\
Err on the side of strictness: if the connection feels forced, don’t go above 6.\\
This is not about keyword overlap — it’s about human intent, curiosity, and conversational flow.\\

Return a JSON object with a top-level key score, which is a list of objects. Each object must include a sentence\_id, a numerical relevance\_score, and a relevance\_score\_reason explaining why that score was given.\\

Summary of Presentation So Far:\\
\{\{Background\_Context\}\}\\

Pervious Sentences:\\
\{\{Prev\_Sentences\}\}\\

Presentation Transcript Segment:\\
\{\{Transcript\_Chunk\}\}\\

Identified Information Needs:\\
\{\{Information\_Need\}\}\\

Strictly follow the below JSON output format\\
Output Format (JSON):\\
\{
  "score": [
    \{
      "sentence\_id": "sentence\_id",
      "relevance\_score\_reason": "reason for the relevance score",
      "relevance\_score": "relevance score"
    \},
    \{
      "sentence\_id": "sentence\_id",
      "relevance\_score\_reason": "reason for the relevance score",
      "relevance\_score": "relevance score"
    \}
  ]
\}
\end{prompt}

\begin{prompt}{End Time Judgement}
You are an expert transcript analyzer. Your task is to determine how long a specific information need remains relevant in a conversation.\\

You will be given the following:\\
- A summary of the transcript up to the current point.\\
- A few sentences before the current transcript segment (Prev\_Sentences).\\
- A few sentences after the current transcript segment (Next\_Sentences).\\
- The current transcript segment being analyzed.\\
- An information need that this segment potentially addresses.\\

Based on the context, identify the last sentence (within the current and next few sentences) where this information need is still relevant. If it's no longer relevant immediately after the segment, return the current segment's last sentence.\\

If there are multiple information needs, you need to find the end\_sentence\_id for each of the needs\\
Only use the given context — do not assume anything beyond what is provided.\\
---\\

Summary:\\
\{\{Summary\}\}\\

Previous Sentences:\\
\{\{Prev\_Sentences\}\}\\

Transcript Segment:\\
\{\{Transcript\_Chunk\}\}\\

Next Sentences:\\
\{\{Next\_Sentences\}\}\\

Information Need:\\
\{\{Information\_Need\}\}\\

---\\

Return a JSON object with a top-level key end\_time, which is a list of objects. Each object corresponds to one information need and includes:\\
- `end\_sentence\_id`: the ID of the last sentence where the need is relevant\\
- `reason`: a brief explanation for why that sentence was chosen\\
Ensure the output is strictly JSON-compliant and follows the below format. \\

Output Format (JSON):\\
\{
 "end\_time": [
        \{
            "end\_sentence\_id": "42",
            "reason": "The discussion about time management ends at this point."
        \},
        \{
            "end\_sentence\_id": "45",
            "reason": "The speaker stops referencing motivation strategies here."
        \}
    ]
\}
\end{prompt}

\begin{prompt}{Question Formatter}
Convert the following sentence into a question. Questions need to be Wh- questions\\

Input:\\

Sentence:\\
\{\{text\}\}\\

Output Format (JSON):\\
\{"question" : "Generated Question"\}\\
\end{prompt}

\section{Detailed Analysis of Human Annotations}
\label{app:human_annotation_analysis}
In this section, we present the results of a detailed analysis of the human-annotated scenes. Our goal is two-fold: First, as JIR is a new task, we would like to better understand this task by examining the variations of human annotations. Due to the variable backgrounds of users, we expected the users to have different needs, but the interesting question here is to what extent their needs vary and 
whether they may also share some common information needs. Second, we are interested in making a comparison between LLM annotations and human annotations. To this end, we have four annotators watch the scenes and record the timestamp and content of questions, if any. 

\begin{figure}[!h]
    \centering
    \includegraphics[width=1.0\linewidth]{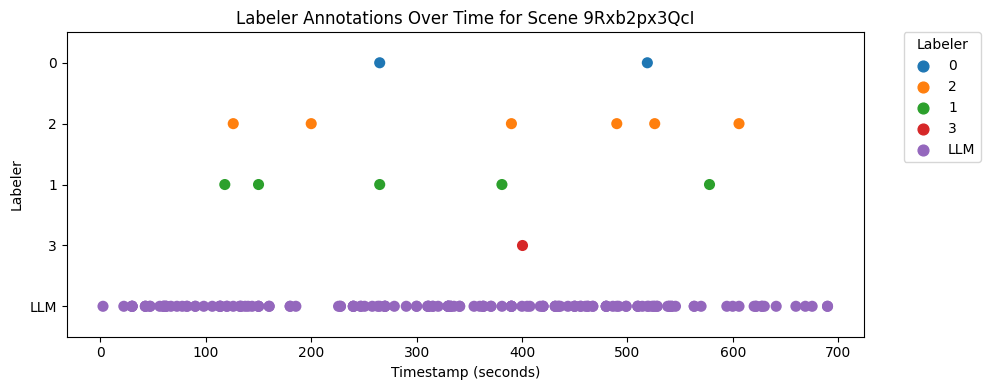}
    \caption{Example of labeled needs across scene 9Rxb2px3Qcl in addition to the generated needs by the language models.}
    \label{fig:scene-distribution}
\end{figure}

In Figure~\ref{fig:scene-distribution}, we show the distribution of needs for a single scene. The x-axis is the timestamp, and the y-axis (and color) indicates the labeler. The graph includes the four annotators along with the LLMs.

We can make several observations from Figure~\ref{fig:scene-distribution}: First, as expected, the needs identified by the human annotators vary significantly in both the number of needs and the timing of the needs. 
We found this to be the case in all the analyzed human-annotated scenes. Using a window size of ten seconds before and after a reference time point,  we can further compute the time-wise ``unique'' human needs - namely, those that occur more than ten seconds away from any other human need. The percentages of such unique needs for all human videos are shown in  Table~\ref{tab:average-unique-human-needs}.
From Table~\ref{tab:average-unique-human-needs}, we see that, in many cases, the majority of human-generated needs were unique time-wise, indicating significant variances across people in the time when a need occurs. However, we should note that in a real application scenario, a deviation of 10 seconds in recommendation of information may not affect the utility of the recommended information that much, so the natural variation of the time stamp among human annotators also means that when we evaluate a JIR system, we may also tolerate a small amount of deviation in the time of recommendation.      

\begin{table}[!h]
    \centering
      {\footnotesize
    \begin{tabular}{|c|c|c|c|c|c|}
    \hline
    Scene ID & 9Rxb2px3QcI & Dxit4BOO0y8 & TU19Orwu4jE & k$\_$gGx9ObXeM & IFKnq9QM6$\_$A\\
    \hline
    Percentage & 42.9\% & 52.7\% & 68.6\% & 54.5\% & 58.8\% \\
    \hline
    \hline
    Scene ID & MXJ-zpJeY3E & ghZRzOb$\_$bZo & v68zYyaEmEA & k$\_$wvXDB9dMdEo & \\
    \hline
    Percentage & 33.3\% & 54.5\% & 87.5\% & 66\% & \\
    \hline
    \end{tabular}}
    \caption{The percentage of time-wise unique human needs for each scene.}
    \label{tab:average-unique-human-needs}
\end{table}

Second, we see that the LLMs identified many more needs than the human annotators. Based on human verification, most of the identified needs by LLMs are reasonable. There may be two reasons why human annotators did not identify so many needs as the LLMs: 1) Humans had limited cognitive processing capacity, and they were only able to identify some of the needs while watching a video in one pass; if human annotators were allowed to re-watch the video multiple times, they might also be able to identify more needs. 2) Humans tend to be satisfied with a ``basic" understanding of the content, whereas the LLMs sometimes ``demanded" more detailed justification or explanation; the LLMs appear to have a tendency to ask for an elaboration whenever a general statement is made (e.g., if A is said to be better than B, the LLM may ask for evidence). From an application perspective, the much larger number of needs identified by the LLMs makes the dataset potentially better represent a wider range of needs of real users than using humans to identify the needs. We have also found that the generated questions by LLMs are of multiple types, including both elaboration questions and why questions. 

\begin{figure}[!h]
    \centering
    \includegraphics[width=1.0\linewidth]{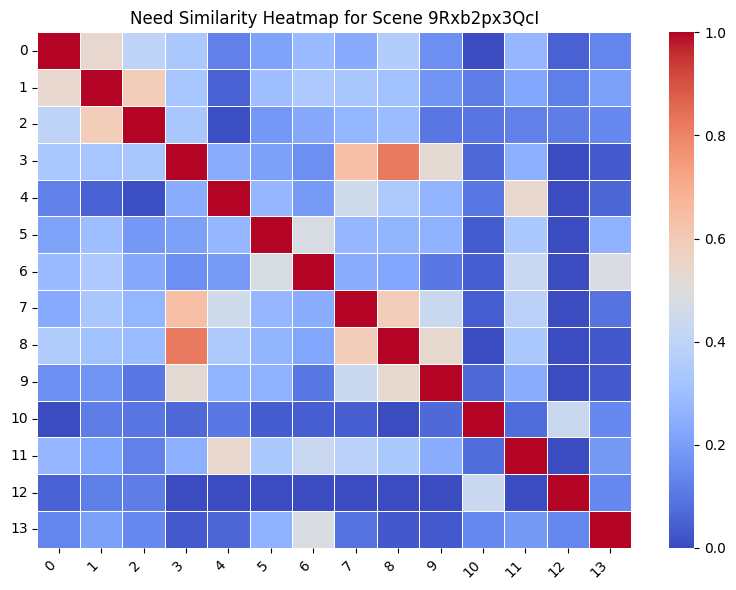}
    \caption{Example of human-generated need similarity for scene 9Rxb2px3Qcl.}
    \label{fig:scene-distribution-heatmap}
\end{figure}

Next, we also compared the semantic similarity of the human-generated needs, both to each other and to the LLM-generated needs. As an example, Figure~\ref{fig:scene-distribution-heatmap} depicts the semantic similarity heatmap comparing all human-generated needs for scene 9Rxb2px3Qcl. 

From the example depicted in Figure~\ref{fig:scene-distribution-heatmap}, and the others, we found that many of the human-generated needs are semantically dissimilar, showing that not only did the time of need vary, but the specific information needed by humans also varies significantly. In Table~\ref{tab:average-unique-human-needs-sim}, we list the mean similarity and standard deviation for all human-generated needs per scene. Similarity was measured using MiniLM-L6-V2 from sentence-transformers~\cite{reimers-2020-Curse_Dense_Retrieval}. The low mean similarity shows that the needs identified from the same video are semantically different. This may be expected to some extent because at a different time point, the content discussed in the scene is likely different and thus the need is also likely different semantically. However, what is somewhat surprising is that even the needs identified in adjacent time windows do not always have a high similarity, again suggesting variable information needs among humans. This also suggests that in a real application, it would be beneficial for a JIR system to allow a user to enter a query as well, which would facilitate human-AI collaboration in precisely specifying the information need. However, we do see that in some time periods (e.g,. at the very beginning and middle of this video), the adjacent distinct needs have higher semantic similarity, forming a small cluster. This suggests that there are also common needs shared by human annotators (those are likely similar needs identified at slightly different time points). Such common needs can also be visually seen from Figure~\ref{fig:scene-distribution}. 

\begin{table}[!h]
    \centering
    {\scriptsize
    \begin{tabular}{|c|c|c|c|c|c|}
    \hline
    Scene ID & 9Rxb2px3QcI & Dxit4BOO0y8 & TU19Orwu4jE & k$\_$gGx9ObXeM & IFKnq9QM6$\_$A\\
    \hline
    Similarity & 0.2195 & 0.1630\% & 0.1896 & 0.1734 & 0.2904 \\
    Std. Dev. & 0.1839 & 0.1668\% & 0.1549 & 0.1493 & 0.2069 \\
    \hline
    \hline
    Scene ID & MXJ-zpJeY3E & ghZRzOb$\_$bZo & v68zYyaEmEA & k$\_$wvXDB9dMdEo & \\
    \hline
    Similarity & 0.1858 & 0.3491\% & 0.1673 & 0.1951 &  \\
    Std. Dev. & 0.1605 & 0.2406\% & 0.1597 & 0.1329 &  \\
    \hline
    \end{tabular}}
    \caption{The mean similarity and standard deviation for human-generated needs of each scene.}
    \label{tab:average-unique-human-needs-sim}
\end{table}

\begin{figure}[!h]
    \centering
    \includegraphics[width=1.0\linewidth]{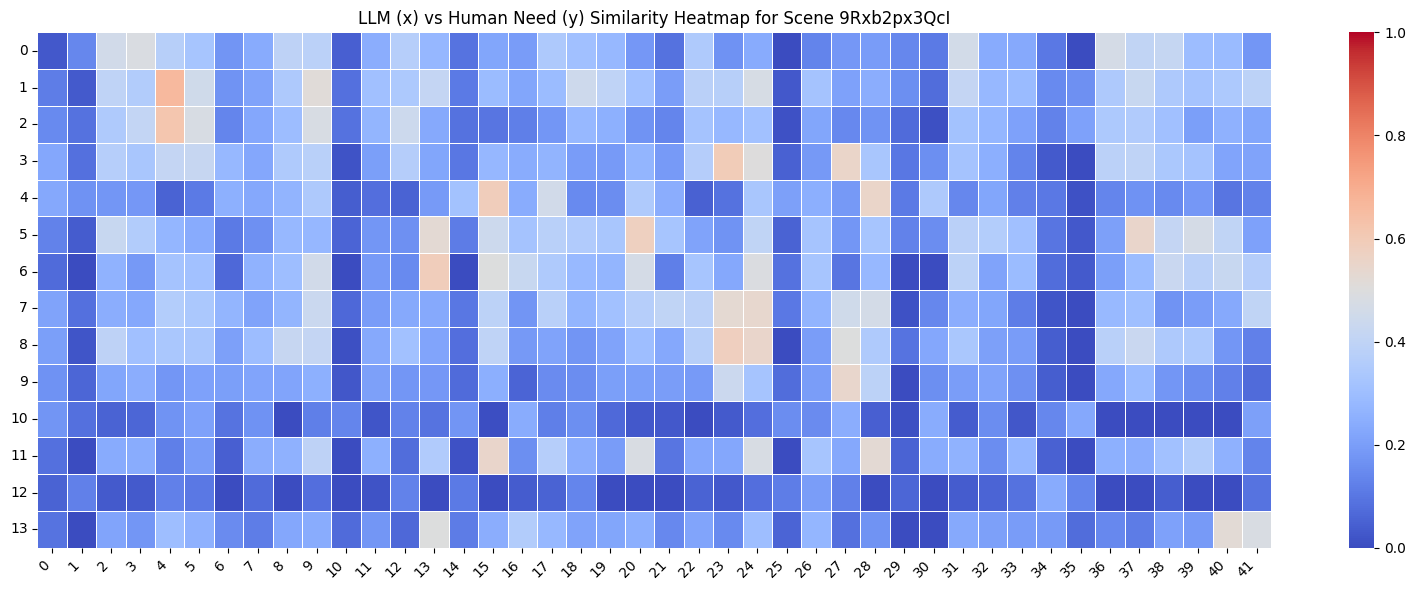}
    \caption{Example of similarity between LLM-generated needs and human-generated needs for scene 9Rxb2px3Qcl.}
    \label{fig:scene-distribution-heatmap-llm}
\end{figure}

Finally, we compared the human annotations to the LLM-generated annotations. To do so, we again applied the window, and only included LLM-generated needs if they fell within ten seconds of a human-generated need. We depict an example of this in Figure~\ref{fig:scene-distribution-heatmap-llm}.

Here, we can see a slightly similar diagonal, indicating that the language models were able to generate needs that overlap with some of the human-generated needs, especially the shared common needs by human annotators at the beginning and in the middle. We further measured the number of human-generated needs that are covered by the language models (namely, above a similarity threshold of 0.55) and we report these results in Table~\ref{tab:covered-human-needs}. The results show that the coverage varies across scenes. A lecture-style scene on a focused topic (i.e., ghZRzOb$\_$bZo  ) may have a coverage as high as over 90\%, while an informal narrative scene about a person (i.e., MXJ-zpJeY3E) may have a coverage as low as 25\%. Intuitively, this is expected, and in general, the coverage is related to the entropy of the information covered in the scene with a higher entropy likely correlated with a low coverage (as it would be intuitively harder to predict the information need). Further investigating this kind of hypothesis by doing more systematic analysis would be an interesting future direction. It also shows that our dataset can also be potentially useful for analysis of topics covered in those scenes, as a ``by product". From the application perspective, a high coverage means that a JIR application system can be expected to serve many users very well via pure recommendation without requiring any user effort in the case of ``low entropy scenes" (predicable information needs), while in the case of ``high entropy scenes" (unpredictable information needs), the system would benefit from providing a query box to a user to accommodate ad hoc variations of their unique needs. Since the entropy can be measured using an algorithm, such an adaptive interface may be even automatically adjusted by an algorithm. This is another useful insight that we were able to derive from human annotations.

\begin{table}[!h]
    \centering
    {\scriptsize
    \begin{tabular}{|c|c|c|c|c|c|}
    \hline
    Scene ID & 9Rxb2px3QcI & Dxit4BOO0y8 & TU19Orwu4jE & k$\_$gGx9ObXeM & IFKnq9QM6$\_$A\\
    \hline
    Percentage & 50\% & 58.3\% & 77.1\% &  59.1\% & 70.6\% \\
    \hline
    \hline
    Scene ID & MXJ-zpJeY3E & ghZRzOb$\_$bZo & v68zYyaEmEA & k$\_$wvXDB9dMdEo & \\
    \hline
    Percentage & 25\% & 90.1\% & 43.8\% & 61.7\% & \\
    \hline
    \end{tabular}}
    \caption{The percentage of human needs covered by language models.}
    \label{tab:covered-human-needs}
\end{table}
\section{LLM Prompts for Information Retrieval Quality Check}
\label{app:ir_prompts}
\begin{prompt}{Information Retrieval Quality Check}
You are an assistant that evaluates whether a given query can be answered using the content of a provided reference or source material.\\
Your task is to:\\

1. Determine whether the reference material contains sufficient information to fully or partially answer the query.\\
2. If yes, extract the relevant parts of the reference that help answer the query.\\
3. Return your answer in the following JSON format:\\

```json\\
\{
  "answerable": true | false,
  "supporting\_content": "If answerable is true, copy and paste only the relevant original text from the reference material that helps answer the query. Do not rephrase, summarize, or explain—only extract directly. If answerable is false, leave this as an empty string."
\}\\
```

\{\{customed\_instruction\}\}\\

Here is the query:\\
\{\{query\}\}\\

Here is the reference material:\\
\{\{reference\_material\}\}\\

Only return the JSON object, with no extra explanation or formatting.
\end{prompt}

\begin{prompt}{Customed Instructions}
Wikipedia: \\
The reference material comes from a Wikipedia article. This article is not the source where the query comes from, so if the query is asking about the specific content, methods, or claims of the paper/study being presented, this reference is insufficient. In such cases, the correct response is \"answerable\": false.\\

Arxiv:\\
The reference material includes the title and abstract of an arXiv paper. This is not the source where the query comes from, so if the query is asking about the specific content, methods, or claims of the paper/study being presented, this reference is insufficient. In such cases, the correct response is \"answerable\": false.\\

Paper:\\
The reference material is extracted from the body of the research paper being presented. It may have been converted from a PDF and could contain distortions such as missing figures with only captions, incomplete tables, or malformed equations. Ignore these potential issues and assume the text reflects the intended content of the original paper. Your task is to evaluate whether the given query can be answered based on the provided excerpt, even partially, as if it were a faithful representation of the original document.\\

Textbook:\\
The reference material comes from a relevant textbook specialized on the topic being discussed.
    
\end{prompt}
\section{Details of the Baseline JIR Systems}
\label{app:baseline}
\subsection{Illustration of the Pipeline of a Prototypical JIR System}
\begin{figure*}[ht]
    \centering
    \includegraphics[trim={0 0.07cm 0 0}, width=1.0\textwidth]{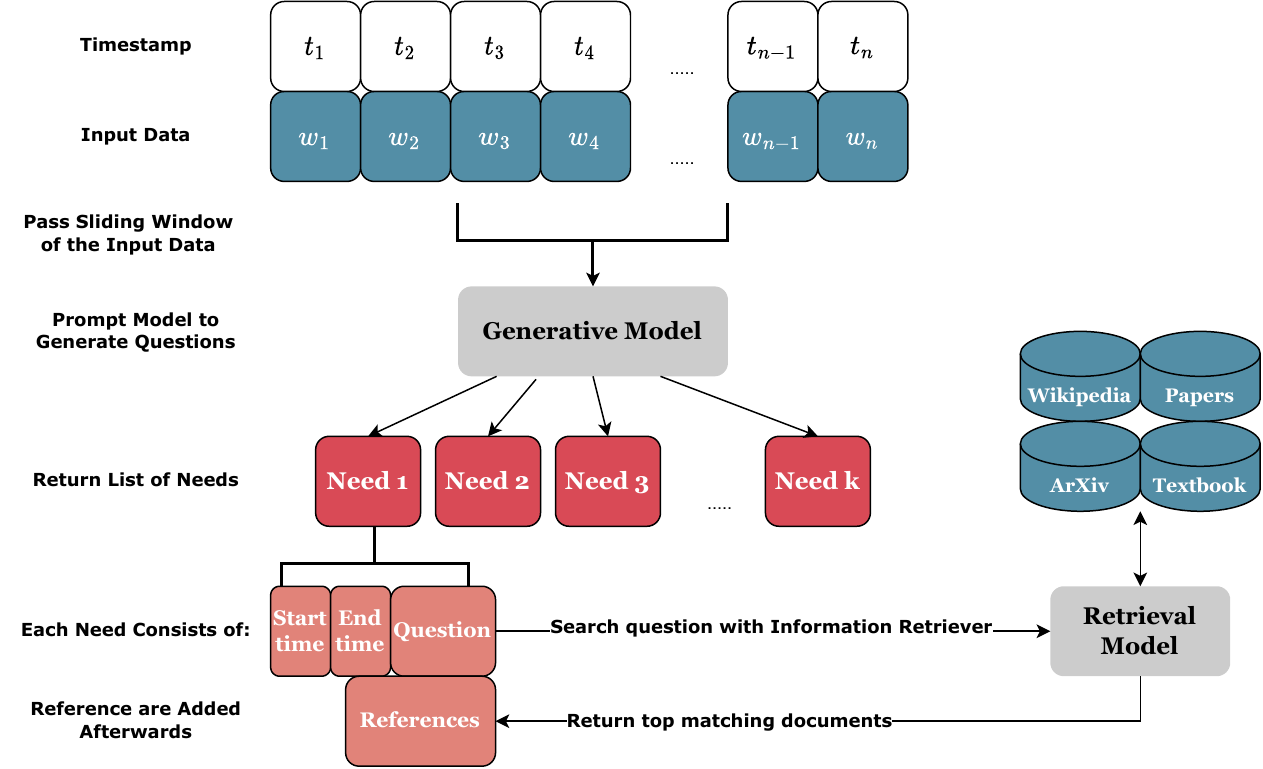}
    \caption{\label{fig:baseline}
    The prototypical JIR system pipeline.
    }
\end{figure*}
This is the naive version of a JIR system, which: \textit{i)} considers a sliding window of the multimodal context, \textit{ii)} generates user information need with a generative model, and \textit{iii)} for each need, retrieves relevant documents to address the need. In our implementation, we consider the most informative text modal input, using generative models \gpt, \ds, \claude, and \gemini, and adopt \textsc{OpenSearch} as our retrieval model.

\subsection{Information Need Detection Prompts}
\begin{prompt}{}
Given the transcript formatted by lines of timestamps/words, predict the questions that a listener might have. Respond in the following JSON format:\\

            {
                "needs": [
                    
                    {
                        "start\_time": <str, the earliest timestamp that the specific question may appear>,
                        "end\_time": <str, the latest timestamp that the specific question may appear>,
                        "question": <str, the description of the question>
                    }
                ]
            
            }
  
            Here is the transcript: \\
            <begin\_transcript>\\
            
            <REPLACE\_WITH\_TRANSCRIPT>\\
            
            <end\_transcript>\\
            
            As a reminder, given the above transcript with timestamps, predict the questions that a listener might have. Respond in the following JSON format:\\

            {
                "needs": [
                    
                    {
                        "start\_time": <str, the earliest timestamp that the specific question may appear>,
                        "end\_time": <str, the latest timestamp that the specific question may appear>,
                        "question": <str, the description of the question>
                    }
                ]
            
            }

            Make sure to follow these rules\\
            - Include all context in a question so that it can be answered without the transcript\\
            - Only include a single question/need per needs block.\\
            - Simulate a curious listener interested in the presented topic - questions should be reasonably likely to be asked by an attendee of this presentation.\\
            
            Respond only in the above JSON format, with nothing else.\\
\end{prompt}
\section{Limitations}\label{limitations}

Despite the promising directions of \jirarena, there are some significant limitations to our work. First, the curated scenes are limited in nature, as only popular learning-based videos (i.e., conference and lecture-style) were selected. Similarly, the retrieval scopes are also scoped, meaning that certain needs may not be answerable from the curated resources. Moreover, the selected scenes were technical, meaning that they might not apply to broader settings where just-in-time recommendation agents may be used, such as enterprise or personal settings. A second limitation is the human-verification, as our manual annotations were inherently limited in scope. In an ideal setting, such verification should be conducted at a broader scale, and should cover the entire benchmark dataset. More broadly, future work may consider adding personalization to such verification, ensuring not just basic need coverage but coverage with respect to varying levels of background knowledge or goals. 

A third limitation is the reliance on closed-source language models. Given their instability, it may be difficult for others to exactly recreate our dataset in the future, as the models may change or be updated. And finally, a fourth limitation is in the methodologies of implementation. Given the lack of a ``ground truth,'' many generated needs are difficult to objectively assess, resulting in the requirement of made assumptions. It may be the case that different settings require different assumptions, or that different assumptions may lead to stronger results in a given setting. Therefore, we encourage future researchers to leverage this as a first step towards understanding this landscape, the necessary assumptions, and how to create and measure JIR agents. 
\section{Broader Impacts}\label{broader impacts}

While it is unlikely that there are direct negative impacts or harms that arise from our proposed benchmark, it is worth considering the potential second-order negative effects of JIR systems. One potential negative effect is the lack of friction and its effect on learning. Namely, individuals may learn better when it is harder to answer a question, or when they have to come up with the question themselves. If a system always provides them the ``right question to ask,'' then they may become reliant on the system for their exploration. Because the overall space of just-in-time recommendation agents is fairly nascent, the effects of these systems on learning or exploration remains an open research question. 



\end{document}